\documentclass[aps,prd,twocolumn,superscriptaddress,showpacs,preprintnumbers,
               amsmath,amssymb]{revtex4-1}

\usepackage{epsfig}

\usepackage{dcolumn}  

\usepackage{color}
\usepackage[mathlines]{lineno}
\usepackage{xspace}
\usepackage{graphicx}
\usepackage{amsmath}
\usepackage{relsize}
\usepackage{float}
\usepackage{array}
\usepackage{tabularx}
\usepackage[dvipsnames]{xcolor}
\usepackage{slashed}
\usepackage[normalem]{ulem}
\graphicspath{{ps}}

\usepackage{hyperref}

\hypersetup{
colorlinks=true,
linkcolor=blue,
citecolor=blue,
urlcolor=red,
linktoc=all
}

\def\babar{\mbox{\slshape B\kern-0.1em{\smaller A}\kern-0.1em
    B\kern-0.1em{\smaller A\kern-0.2em R}}\xspace}
\def\bztoltau {\ensuremath{B^{0} \to \tau^\pm \ell^\mp}\xspace}
\def\boldbztoltau {\boldmath \ensuremath{B^{0} \to \tau^\pm \ell^\mp}\xspace}
\def\bztomutau {\ensuremath {B^{0} \to \tau^\pm \mu^{\mp}}\xspace}
\def\bztoetau {\ensuremath {B^{0} \to \tau^\pm e^{\mp}}\xspace}

\def\bztodpi {\ensuremath {B^{0}\to D^{(*)-} \pi^+}\xspace}
\def\bztodpiplus {\ensuremath {B^{0}\to D^{-} \pi^+}\xspace}
\def\bztodstarpi {\ensuremath {B^{0}\to D^{*-} \pi^+}\xspace}

\begin{document}

\title{Search for \boldbztoltau ($\ell=e,\,\mu$)
with a hadronic tagging method at Belle}


\affiliation{Department of Physics, University of the Basque Country UPV/EHU, 48080 Bilbao}
\affiliation{University of Bonn, 53115 Bonn}
\affiliation{Brookhaven National Laboratory, Upton, New York 11973}
\affiliation{Budker Institute of Nuclear Physics SB RAS, Novosibirsk 630090}
\affiliation{Faculty of Mathematics and Physics, Charles University, 121 16 Prague}
\affiliation{Chonnam National University, Gwangju 61186}
\affiliation{University of Cincinnati, Cincinnati, Ohio 45221}
\affiliation{Deutsches Electronen-Synchrotron, 22607 Hamburg}
\affiliation{University of Florida, Gainesville, Florida 32611}
\affiliation{Department of Physics, Fu Jen Catholic University, Taipei 24205}
\affiliation{Key Laboratory of Nuclear Physics and Ion-beam Application (MOE) and Institute of Modern Physics, Fudan University, Shanghai 200443}
\affiliation{Justus-Liebig-Universit\"at Gie\ss{}en, 35392 Gie\ss{}en}
\affiliation{Gifu University, Gifu 501-1193}
\affiliation{II. Physikalisches Institut, Georg-August-Universit\"at G\"ottingen, 37073 G\"ottingen}
\affiliation{SOKENDAI (The Graduate University for Advanced Studies), Hayama 240-0193}
\affiliation{Gyeongsang National University, Jinju 52828}
\affiliation{Department of Physics and Institute of Natural Sciences, Hanyang University, Seoul 04763}
\affiliation{University of Hawaii, Honolulu, Hawaii 96822}
\affiliation{High Energy Accelerator Research Organization (KEK), Tsukuba 305-0801}
\affiliation{National Research University Higher School of Economics, Moscow 101000}
\affiliation{Forschungszentrum J\"{u}lich, 52425 J\"{u}lich}
\affiliation{IKERBASQUE, Basque Foundation for Science, 48013 Bilbao}
\affiliation{Indian Institute of Science Education and Research Mohali, SAS Nagar, 140306}
\affiliation{Indian Institute of Technology Bhubaneswar, Satya Nagar 751007}
\affiliation{Indian Institute of Technology Guwahati, Assam 781039}
\affiliation{Indian Institute of Technology Hyderabad, Telangana 502285}
\affiliation{Indian Institute of Technology Madras, Chennai 600036}
\affiliation{Indiana University, Bloomington, Indiana 47408}
\affiliation{Institute of High Energy Physics, Chinese Academy of Sciences, Beijing 100049}
\affiliation{Institute of High Energy Physics, Vienna 1050}
\affiliation{Institute for High Energy Physics, Protvino 142281}
\affiliation{INFN - Sezione di Napoli, I-80126 Napoli}
\affiliation{INFN - Sezione di Roma Tre, I-00146 Roma}
\affiliation{INFN - Sezione di Torino, I-10125 Torino}
\affiliation{Advanced Science Research Center, Japan Atomic Energy Agency, Naka 319-1195}
\affiliation{J. Stefan Institute, 1000 Ljubljana}
\affiliation{Institut f\"ur Experimentelle Teilchenphysik, Karlsruher Institut f\"ur Technologie, 76131 Karlsruhe}
\affiliation{Kavli Institute for the Physics and Mathematics of the Universe (WPI), University of Tokyo, Kashiwa 277-8583}
\affiliation{Kennesaw State University, Kennesaw, Georgia 30144}
\affiliation{Department of Physics, Faculty of Science, King Abdulaziz University, Jeddah 21589}
\affiliation{Kitasato University, Sagamihara 252-0373}
\affiliation{Korea Institute of Science and Technology Information, Daejeon 34141}
\affiliation{Korea University, Seoul 02841}
\affiliation{Kyoto Sangyo University, Kyoto 603-8555}
\affiliation{Kyungpook National University, Daegu 41566}
\affiliation{Universit\'{e} Paris-Saclay, CNRS/IN2P3, IJCLab, 91405 Orsay}
\affiliation{P.N. Lebedev Physical Institute of the Russian Academy of Sciences, Moscow 119991}
\affiliation{Faculty of Mathematics and Physics, University of Ljubljana, 1000 Ljubljana}
\affiliation{Ludwig Maximilians University, 80539 Munich}
\affiliation{Luther College, Decorah, Iowa 52101}
\affiliation{Malaviya National Institute of Technology Jaipur, Jaipur 302017}
\affiliation{Faculty of Chemistry and Chemical Engineering, University of Maribor, 2000 Maribor}
\affiliation{Max-Planck-Institut f\"ur Physik, 80805 M\"unchen}
\affiliation{School of Physics, University of Melbourne, Victoria 3010}
\affiliation{University of Mississippi, University, Mississippi 38677}
\affiliation{University of Miyazaki, Miyazaki 889-2192}
\affiliation{Moscow Physical Engineering Institute, Moscow 115409}
\affiliation{Graduate School of Science, Nagoya University, Nagoya 464-8602}
\affiliation{Universit\`{a} di Napoli Federico II, I-80126 Napoli}
\affiliation{Nara Women's University, Nara 630-8506}
\affiliation{National Central University, Chung-li 32054}
\affiliation{National United University, Miao Li 36003}
\affiliation{Department of Physics, National Taiwan University, Taipei 10617}
\affiliation{H. Niewodniczanski Institute of Nuclear Physics, Krakow 31-342}
\affiliation{Nippon Dental University, Niigata 951-8580}
\affiliation{Niigata University, Niigata 950-2181}
\affiliation{Novosibirsk State University, Novosibirsk 630090}
\affiliation{Osaka City University, Osaka 558-8585}
\affiliation{Pacific Northwest National Laboratory, Richland, Washington 99352}
\affiliation{Panjab University, Chandigarh 160014}
\affiliation{Peking University, Beijing 100871}
\affiliation{University of Pittsburgh, Pittsburgh, Pennsylvania 15260}
\affiliation{Research Center for Nuclear Physics, Osaka University, Osaka 567-0047}
\affiliation{Dipartimento di Matematica e Fisica, Universit\`{a} di Roma Tre, I-00146 Roma}
\affiliation{Department of Modern Physics and State Key Laboratory of Particle Detection and Electronics, University of Science and Technology of China, Hefei 230026}
\affiliation{Seoul National University, Seoul 08826}
\affiliation{Soongsil University, Seoul 06978}
\affiliation{Universit\'{e} de Strasbourg, CNRS, IPHC, UMR 7178, 67037 Strasbourg}
\affiliation{Sungkyunkwan University, Suwon 16419}
\affiliation{School of Physics, University of Sydney, New South Wales 2006}
\affiliation{Department of Physics, Faculty of Science, University of Tabuk, Tabuk 71451}
\affiliation{Tata Institute of Fundamental Research, Mumbai 400005}
\affiliation{Department of Physics, Technische Universit\"at M\"unchen, 85748 Garching}
\affiliation{School of Physics and Astronomy, Tel Aviv University, Tel Aviv 69978}
\affiliation{Toho University, Funabashi 274-8510}
\affiliation{Department of Physics, Tohoku University, Sendai 980-8578}
\affiliation{Earthquake Research Institute, University of Tokyo, Tokyo 113-0032}
\affiliation{Department of Physics, University of Tokyo, Tokyo 113-0033}
\affiliation{Tokyo Institute of Technology, Tokyo 152-8550}
\affiliation{Tokyo Metropolitan University, Tokyo 192-0397}
\affiliation{Virginia Polytechnic Institute and State University, Blacksburg, Virginia 24061}
\affiliation{Wayne State University, Detroit, Michigan 48202}
\affiliation{Yamagata University, Yamagata 990-8560}
\affiliation{Yonsei University, Seoul 03722}
  \author{H.~Atmacan}\affiliation{University of Cincinnati, Cincinnati, Ohio 45221} 
  \author{A.~J.~Schwartz}\affiliation{University of Cincinnati, Cincinnati, Ohio 45221} 
  \author{K.~Kinoshita}\affiliation{University of Cincinnati, Cincinnati, Ohio 45221} 
  \author{I.~Adachi}\affiliation{High Energy Accelerator Research Organization (KEK), Tsukuba 305-0801}\affiliation{SOKENDAI (The Graduate University for Advanced Studies), Hayama 240-0193} 
  \author{K.~Adamczyk}\affiliation{H. Niewodniczanski Institute of Nuclear Physics, Krakow 31-342} 
  \author{H.~Aihara}\affiliation{Department of Physics, University of Tokyo, Tokyo 113-0033} 
  \author{S.~Al~Said}\affiliation{Department of Physics, Faculty of Science, University of Tabuk, Tabuk 71451}\affiliation{Department of Physics, Faculty of Science, King Abdulaziz University, Jeddah 21589} 
  \author{D.~M.~Asner}\affiliation{Brookhaven National Laboratory, Upton, New York 11973} 
  \author{V.~Aulchenko}\affiliation{Budker Institute of Nuclear Physics SB RAS, Novosibirsk 630090}\affiliation{Novosibirsk State University, Novosibirsk 630090} 
  \author{T.~Aushev}\affiliation{National Research University Higher School of Economics, Moscow 101000} 
% ...
\collaboration{The Belle Collaboration}


\begin{abstract}

 We present a search for the lepton-flavor-violating decays \bztoltau,
 where $\ell=(e,\,\mu)$, using the full data sample of
 $772 \times 10^6$ $B \overline{B}$ pairs
 recorded by the Belle
 detector at the KEKB asymmetric-energy $e^+e^-$ collider.
 We use events in which one $B$ meson
 is fully reconstructed in a hadronic decay mode. The $\tau^\pm$
 lepton is reconstructed
 indirectly using the momentum of the reconstructed $B$ and
 that of the $\ell^\mp$ from the signal decay.
 We find no evidence for \bztoltau decays and set
 upper limits on their branching fractions at 90\% confidence level
 of  {$\cal B$}(\bztomutau)~$< 1.5 \times 10^{-5}$ and
 {$\cal B$}(\bztoetau)~$< 1.6 \times 10^{-5}$.

\end{abstract}

\pacs{13.25.Hw, 14.40.Nd}

\maketitle


\section{INTRODUCTION}

 The lepton-flavor-violating decays \bztoltau~\cite{chargeconj},
 where $\ell= (e,\,\mu)$, are 
 promising modes in which to search for new physics.
 Recently, there have been indications of 
 possible violation of lepton flavor universality (LFU) in
 $B^0\to D^{(*)-}\tau^+\nu$~\cite{hflavR_D},
 $B^0\to K^{*0}\ell^+\ell^-$~\cite{lhcbR_Kstar} and $B^\pm\to K^\pm\ell^+\ell^-$~\cite{lhcbR_K1,lhcbR_K2} decays.
 Other studies are less conclusive~\cite{belleRKstar, babarRKstar}.
 LFU violation is often accompanied by 
 lepton flavor violation (LFV) in theoretical models~\cite{theoLFULFV}.
 The decay \bztoltau, like $B^0\to D^{(*)-}\tau^+\nu$,
 connects a third-generation quark with a
 third-generation lepton.
 The decay can occur in principle
 via neutrino mixing~\cite{neutrinooscillation};
 however, the rate due to such mixing~\cite{footnote} is considerably
 below current or future experimental sensitivities.
 Thus, observing these decays would indicate new physics.
 Some new physics models give rise to
 branching fractions of 10$^{-9}$ to 10$^{-10}$.
 For example, Pati-Salam vector leptoquarks of mass
 86~{\rm TeV}/$c^2$ give branching fractions of
 $4.4\times 10^{-9}$ for \bztomutau
 and $1.6 \times 10^{-9}$ for \bztoetau~\cite{vectorleptoquark}.
 The general flavor-universal minimal supersymmetric Standard Model
 predicts branching fractions of up to about
 $2\times 10^{-10}$~\cite{btoltauMSSM}.

 These decay modes have previously been studied by the
 CLEO~\cite{btoltaucleo}, \babar~\cite{btoltauBabar},
 and LHCb~\cite{btoltauLHCb} experiments.
 No evidence for these decays has been found.
 The current most stringent upper limits are
 ${\cal B}(\bztomutau)< 1.2\times 10^{-5}$~\cite{btoltauLHCb} and
 ${\cal B}(\bztoetau) < 2.8\times 10^{-5}$~\cite{btoltauBabar},
 both at 90\% confidence level (CL).
 In this paper we report a search for \bztoltau decays
 using the full Belle data sample of 711~fb$^{-1}$ recorded
 at the $\Upsilon$(4S) resonance.
 This is the first such search from Belle.


\section{DATASET AND DETECTOR DESCRIPTION}

 Our data sample consists of $(772 \pm 11) \times 10^6$
 $B\overline{B}$ pairs produced in $e^+ e^-\to\Upsilon(4S)$ events
 recorded by the Belle detector at the KEKB
 asymmetric-energy $e^+e^-$ collider~\cite{KEKB}.
 The Belle detector is a large-solid-angle magnetic spectrometer
 that consists of a silicon vertex detector (SVD), a 50-layer
 central drift chamber (CDC), an array of aerogel threshold
 Cherenkov counters (ACC), a barrel-like arrangement of
 time-of-flight (TOF) scintillation
 counters, and an electromagnetic calorimeter comprising
 CsI(Tl) crystals (ECL).
 All these detectors are located inside a superconducting
 solenoid coil that provides a 1.5 T magnetic field.
 An iron flux return yoke located outside the coil
 is instrumented with resistive-plate chambers (KLM)
 to detect $K^0_L$ mesons and to identify muons.
 Two inner detector configurations were used:
 for the first $152 \times 10^6$ $B \overline{B}$ pairs,
 a 2.0~cm radius beam-pipe and a three-layer SVD were used;
 and for the remaining $620 \times 10^6$ $B \overline{B}$ pairs,
 a 1.5~cm radius beam-pipe, a four-layer SVD~\cite{SVD},
 and a small-cell inner drift chamber were used.
 A more detailed description of
 the detector is provided in Ref.~\cite{Belle}.
 
 We study properties of signal events, sources of background,
 and optimize selection criteria using Monte Carlo (MC)
 simulated events.
 These samples are generated using the software packages
 {\sc EvtGen}~\cite{evtgen} and {\sc Pythia}~\cite{pythia},
 and final-state radiation is included via {\sc Photos}~\cite{photos}.
 The detector response is simulated using {\sc Geant3}~\cite{geant}.
 We produce \bztoltau MC events to calculate signal
 reconstruction efficiencies.
 To estimate backgrounds, we use MC samples that
 describe all $e^+e^-\to q\bar{q}$ processes.
 Events containing $e^+e^-\to B\overline{B}$
 with subsequent
 $b\to cW$ decay, and $e^+e^- \to q\bar q$ ($q=u,d,s,c$) continuum
 events, are both simulated
 with five times the integrated luminosity of Belle.
 Semileptonic $b \to u \ell \nu$ decays are simulated
 with 20 times the integrated luminosity.
 Rare $b\to s$ and $b\to u$ decays are simulated with
 50 times the integrated luminosity.


\section{EVENT SELECTION}

 Our analysis uses a technique uniquely suited to $e^+e^-$
 flavor factory experiments, in which the energy and momentum
 of the initial state are known.
 We first reconstruct a $B$ meson decaying hadronically;
 this is referred to as the ``tag-side" $B$ meson ($B_{\rm tag}$).
 We use the reconstructed $B_{\rm tag}$ momentum and the $e^+e^-$
 initial momentum to infer the momentum of the signal-side
 $B$ meson ($B_{\rm sig}$).
 Because \bztoltau are two-body decays, the
 momentum of the $\tau$ lepton can be inferred from
 the momentum of $B_{\rm sig}$ and the momentum of $\ell^\mp$;
 thus, the $\tau^\pm$ does not need to be reconstructed.
 We define the ``missing mass" as
 \begin{linenomath*}
  \begin{equation}
   M_{\rm miss}  = \sqrt{(E^{}_{B_{\rm sig}} - E^{}_{\ell})^2/c^4 -
   (\vec{p}^{}_{B_{\rm sig}} - \vec{p}^{}_{\ell})^2/c^2}\,,
  \label{eq:eq1}
  \end{equation}
 \end{linenomath*}
 where $E^{}_{B_{\rm sig}}$ and $\vec{p}^{}_{B_{\rm sig}}$ are the
 energy and momentum, respectively, of $B^{}_{\rm sig}$, and
 $E_{\ell}$ and $\vec{p}^{}_{\ell}$ are the corresponding
 quantities for $\ell^\mp$. The quantity $M^{}_{\rm miss}$ is
 the invariant mass of the unreconstructed or missing particle and,
 for \bztoltau decays, should peak at the mass of the $\tau$ lepton
 ($m^{}_\tau = 1.776$~GeV/$c^2$~\cite{pdg}).
 To improve the resolution in $M^{}_{\rm miss}$, we
 evaluate it in the $e^+e^-$ center-of-mass (c.m.) frame
 and substitute the beam energy $E^{}_{\rm beam}$ for
 $E^{}_{B_{\rm sig}}$.
 To avoid introducing bias in our analysis, we
 analyze the data in a ``blind" manner, i.e, we finalize all selection
 criteria before viewing events in a region around~$m^{}_\tau$.
 This blinded region is $[1.65, 1.90]$~GeV/$c^2$, which
 corresponds to approximately $3.8\sigma$ in the resolution.
 
 \subsection{Tag-side selection}

 We first reconstruct $B^{}_{\rm tag}$ candidates in one of 1104
 hadronic decay channels using a hierarchical algorithm based on
 the NeuroBayes neural network package~\cite{bellefullrecon}.
 The quality of $B_{\rm tag}$ is represented by a single
 classifier output, $O_{\rm NN}$, which ranges from
 0 (background-like) to 1 (signal-like).
 The output $O^{}_{\rm NN}$ is mainly
 determined by the $B_{\rm tag}$ reconstruction. It includes
 event-shape information and significantly suppresses
 $e^+e^- \to q\bar q$ continuum events.
 In addition to $O_{\rm NN}$, two other variables
 are used for selecting $B_{\rm tag}$ candidates:
 the energy difference
 $\Delta E \equiv E_{B_{\rm tag}} - E_{\rm beam}$,
 and the beam-energy-constrained mass
 $M_{\rm bc} \equiv \sqrt{E^2_{\rm beam}/c^4-
 |\vec{p}^{}_{B_{\rm tag}}|^2/c^2}$,
 where
 $E_{B_{\rm tag}}$ and $\vec{p}_{B_{\rm tag}}$ are
 the reconstructed energy and momentum, respectively,
 of~$B_{\rm tag}$. These quantities are evaluated
 in the $e^+ e^-$ c.m. system.
 The $B_{\rm tag}$ candidate is required to satisfy
 $|\Delta E|< 0.05$~GeV. For each signal mode, we choose selection
 criteria on $O_{\rm NN}$ and $M_{\rm bc}$ by optimizing a
 figure-of-merit (FOM).
 The FOM is defined as $\varepsilon_{\rm MC} /\sqrt{N_{\rm B}}$,
 where $\varepsilon_{\rm MC}$ is the reconstruction efficiency
 of signal events as determined from MC simulation, and
 $N_{\rm B}$ is the number of background events expected
 within the signal region $M^{}_{\rm miss}\in [1.65,~1.90]$~GeV/$c^2$.
 Based on FOM studies, we require
 $O_{\rm NN}> 0.082$ for \bztomutau,
 $O_{\rm NN}> 0.095$ for \bztoetau, and
 $M_{\rm bc}>$5.272 GeV/$c^2$ for both modes.

 After all $B_{\rm tag}$ selection criteria are applied,
 about 10\% of \bztomutau events and 8\% of \bztoetau events
 have multiple $B^{}_{\rm tag}$ candidates.
 For such events, we select a single $B^{}_{\rm tag}$
 by choosing the candidate with the highest
 value of~$O_{\rm NN}$. This criterion selects the correct candidate
 90\% of the time, according to MC simulation.

 \subsection{Signal-side selection}

 To reconstruct the signal side, only tracks not
 associated with $B_{\rm tag}$ are considered.
 Such tracks are required to originate
 from the interaction point (IP) 
 and have an impact parameter $|dz|<4.0$~cm along the
 $z$ axis, which points opposite the $e^+$ beam direction.
 We also require $dr<2.0$~cm in the $x$-$y$ plane
 (transverse to the $e^+$  beam direction),
 where $dr = \sqrt{dx^2 + dy^2}$.

 Charged tracks are identified by combining information
 from various subdetectors into a likelihood function
 {$\cal L$}$_{i}$, where $i$~=~$e$, $\mu$,
 $\pi$, $K$, or $p$~\cite{likelihood}.
 Muon candidates are identified based on the response
 of the CDC and KLM~\cite{muID}.
 A track with a likelihood ratio
 ${\cal R}^{}_{\mu} = {\cal L}^{}_{\mu}/
 ({\cal L}^{}_{\mu} + {\cal L}^{}_{\pi} + {\cal L}^{}_{K}) > 0.90$
 is identified as a muon.
 The detection efficiency of this requirement
 is about~89\%, and the pion
 misidentification rate is about 2\%.
 Electron candidates are identified mainly using the ratio of the energy
 deposited in the ECL to the track momentum, the shower shape
 in the ECL, and the energy loss in the CDC.
 A track with a likelihood ratio
 ${\cal R}^{}_{e} =
 {\cal L}^{}_{e}/
 ({\cal L}^{}_{e} +{\cal L}^{}_{\rm hadrons})$~$>0.90$
 is identified as an electron, where ${\cal L}^{}_{\rm hadrons}$
 is the product of probability density functions (PDFs) for
 hadrons~\cite{elecID}.
 The efficiency of this requirement is about~94\%,
 and the pion misidentification rate is about 0.3\%.
 We recover electron energy lost due to bremsstrahlung
 by searching for photons within a cone of radius 50 mrad centered
 around the electron momentum. If such a photon is found, its
 four-momentum (assuming it originated at the IP)
 is added to that of the electron.

 We require that $M_{\rm miss}$ be in the range
 1.40 to 2.20~GeV/$c^2$.
 Every muon or electron candidate satisfying
 this requirement is treated as
 a $B_{\rm sig}$ candidate. After these
 selections, we find that less than
 1\% of \bztomutau and \bztoetau events have multiple
 $B^{}_{\rm sig}$ candidates. These fractions
 are consistent with those from MC simulations.
 For such events, in order to preserve efficiency,
 we retain all such candidates, i.e., we do not apply a
 best-candidate selection.

 \subsection{Background}

 After applying all selection criteria,
 a small amount of background remains. This background
 is studied using MC simulation and found to originate
 mainly from $b\to cW$ and
 $b \to u \ell \nu$ decays. These backgrounds are smoothly falling in
 the $M^{}_{\rm miss}$ distribution.
 However, for \bztomutau candidates, two small peaks
 are observed:
 one at $M_{\rm miss}\approx 1.869$~GeV/$c^2$ and
 the other at $M_{\rm miss}\approx 2.010$~GeV/$c^2$.
 The former corresponds to \bztodpiplus
 decays, while the latter corresponds to \bztodstarpi decays,
 where in both cases the $\pi^+$ is misidentified
 as~$\mu^+$.
 These $B^0\to D^{(*)-}\pi^+$ decays
 are taken into account when fitting the $M^{}_{\rm miss}$
 distribution for the signal yield (described below).

 \subsection{Control samples}

 We use control samples of \bztodpi decays
 to determine corrections to the shapes of the \bztoltau PDFs
 used to fit for the signal yields
 (see Sec.~\ref{sec:maxlikfit}).
 To identify \bztodpi decays, we select pions on
 the signal side rather than leptons.
 Pion candidates are identified using
 $dE/dx$ measured in the CDC, time-of-flight information
 from the TOF, and the photon yield in the ACC.
 A track with a likelihood ratio
 ${\cal R}^{}_{\pi} = {\cal L}^{}_{\pi}/
 ({\cal L}^{}_{\pi} + {\cal L}^{}_{K})
 > 0.90$ is identified as a pion~\cite{likelihood}.
 All other selection criteria are the same as for the \bztoltau
 search. In addition, we veto leptons by requiring
 that ${\cal R}^{}_{\mu} < 0.90$ and ${\cal R}^{}_{e} < 0.90$.
 With the above selection, the pion
 identification efficiency is about 95\%, and the kaon
 misidentification rate is about 5\%.


 \section{MAXIMUM LIKELIHOOD FITS}
 \label{sec:maxlikfit}

 We determine the \bztoltau signal yields by performing
 an unbinned extended maximum-likelihood fit to the
 $M_{\rm miss}$ distributions.
 The PDF used to model
 correctly reconstructed signal decays is a double
 Gaussian for \bztomutau and the sum of three
 Gaussians for \bztoetau.
 These Gaussians are allowed to have different means.
 We also model misreconstructed signal decays
 in which the lepton selected is subject to
 final-state radiation or is not a direct
 daughter in the two-body \bztoltau decay,
 i.e., it originates from $\tau^\pm\to\ell^\pm\nu\bar{\nu}$ or
 $\tau^\pm\to\pi^\pm(\to\ell^\pm\nu)\bar{\nu}$.
 This component is referred to as a ``self-cross-feed"
 signal, and we model it with a
 double Gaussian and an exponential function.
 The fractions of self-cross-feed signal are fixed
 to the values obtained from MC simulation:
 $(5.0\pm 0.2)$\% for \bztomutau
 and $(14.0\pm 0.3)$\% for \bztoetau.
 The self-cross-feed fraction 
 is larger for the electron channel due to a larger
 contribution from $B^0\to\tau^{\pm} e^{\mp}\gamma$ decays.

 The shape parameters of the signal PDFs are obtained
 from MC simulations. We make corrections to these to
 account for small differences observed
 between MC simulation and data. We obtain these correction
 factors by fitting the $M_{\rm miss}$ distributions
 of the high-statistics \bztodpi control samples.
 For the \bztodpi samples, we fit both data and MC events
 and record small shifts observed in the means of the PDFs,
 and nominal differences in the widths.
 We apply these shifts for the means and scaling factors for
 the widths to the \bztoltau signal PDFs.
 The uncertainties in these correction factors are
 accounted for when evaluating systematic uncertainties.

 Background PDFs of all modes are modeled with
 exponential functions.
 The shape parameters for these background PDFs are all
 floated, along with the 
 background and signal yields.
 The PDFs for misidentified \bztodpiplus and \bztodstarpi
 decays are taken to be a double Gaussian and the sum of three
 Gaussians, respectively.

 We validate our fitting procedure and check for fit bias
 using MC simulations. We generate large ensembles of simulated
 experiments, in which the $M^{}_{\rm miss}$ distributions are
 generated from the PDFs used for fitting. We fit
 these ensembles and find that the fitted signal yields
 are consistent with the input values; the mean difference
 is $-0.08\pm 0.05$ events for \bztomutau and $-0.00\pm 0.05$ events
 for \bztoetau. We include these small potential biases when
 evaluating systematic uncertainties. We also fit ensembles of
 fully simulated events and again find the signal yields to be
 consistent with the input values.

\begin{figure}[ht]
  \centering
  \epsfig{file=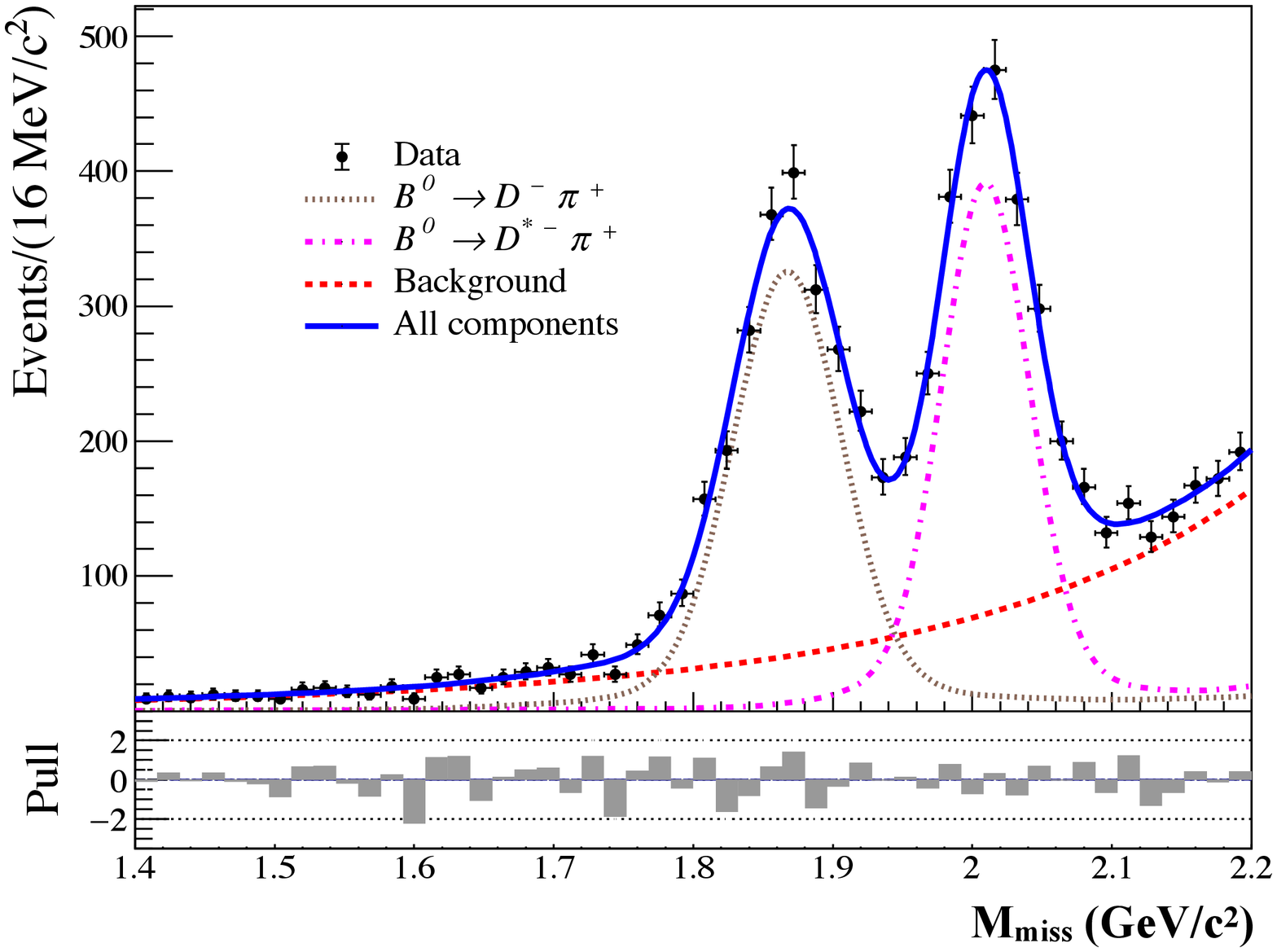,width=1.0\linewidth}
  \caption{
   The $M_{\rm miss}$ distribution of \bztodpi candidates observed
   in data (black dots) along with projections of the fit result:
   the overall fit result (solid blue curve), the background component
   (dashed red curve), the \bztodpiplus component (dotted brown curve)
   and the \bztodstarpi component (dash-dotted magenta curve).
   The plot below the distribution shows the residuals
   divided by the errors (pulls).}
  \label{fig:btodpimlfit}
 \end{figure}

 To further check our analysis procedure and acceptance
 calculation, we measure the branching fractions for the
 control channels \bztodpi.
 The $M^{}_{\rm miss}$ distributions
 of these decays along with projections of the fit result
 are shown in Fig.~\ref{fig:btodpimlfit}.
 To assess the goodness of fit, we calculate
 a $\chi^2$ statistic from the residuals of the fit result.
 Dividing by the number of degrees of freedom ($n^{}_{\rm dof}$)
 gives $\chi^2/n^{}_{\rm dof} = 0.89$, where
 $n^{}_{\rm dof}$ is 41.
 %
 The fitted yields are $2136 \pm 71$ and $2071 \pm 74$ for
 \bztodpiplus and \bztodstarpi, respectively,
 and the resulting branching fractions are
 ${\cal B}(\mbox{\bztodpiplus})=(2.54\pm 0.11)\times 10^{-3}$
 and
 ${\cal B}(\mbox{\bztodstarpi})=(2.67\pm 0.12)\times 10^{-3}$,
 where the uncertainties listed are statistical only. These
 values are in excellent agreement with the current
 world averages
 ${\cal B}(\mbox{\bztodpiplus})=(2.52\pm 0.13)\times 10^{-3}$
 and 
 ${\cal B}(\mbox{\bztodstarpi})= (2.74\pm 0.12)\times 10^{-3}$~\cite{pdg}.

\begin{figure}[ht]
  \centering
   \epsfig{file=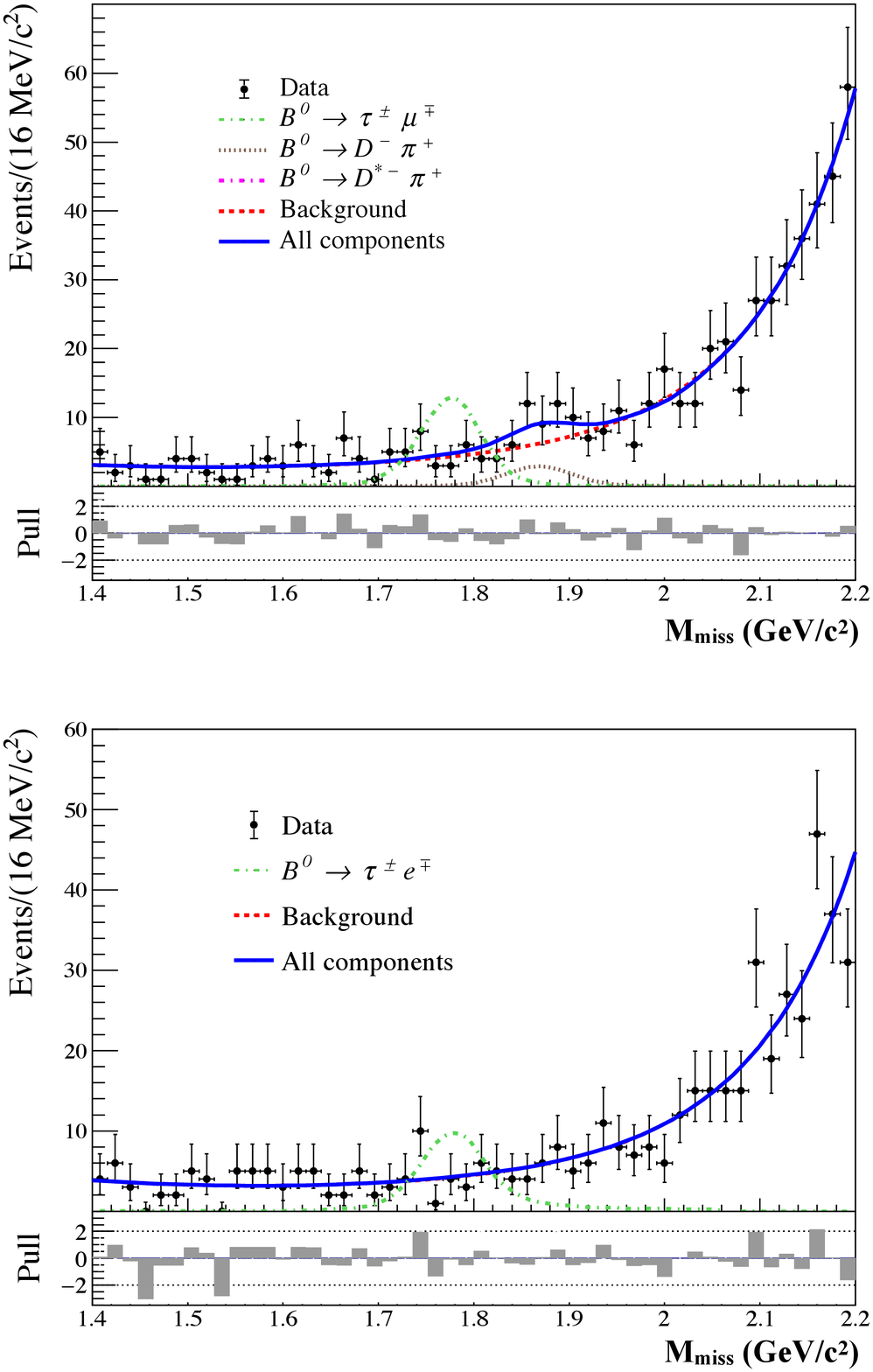,width=1.0\linewidth}
   \caption{
    The $M_{\rm miss}$ distributions of
    \bztomutau (upper) and
    \bztoetau (lower) candidates,
    along with projections of the fit result.
    The black dots show the data, the dashed
    red curve shows the background component,
    and the solid blue curve shows the overall fit result.
    The dash-dotted green curve shows 
    the signal PDF,
    with a normalization
    corresponding to a branching fraction of~$10^{-4}$.
    In the upper plot, the dotted
    brown curve shows the \bztodpiplus component. The plots
    below the distributions show the residuals divided by
    the errors (pulls).}
   \label{fig:muemlfit}
 \end{figure}

 The $M^{}_{\rm miss}$ distributions for signal
 \bztoltau decays along with projections of the fit
 result are shown in Fig.~\ref{fig:muemlfit}.
 The $\chi^2/n^{}_{\rm dof}$ values
 are 0.54 ($n^{}_{\rm dof}$~=~44) and 0.70
 ($n^{}_{\rm dof}$~=~44) for \bztomutau and
 \bztoetau, respectively.
 The fitted signal yields are
 $N^{}_{\rm sig} = 1.8\,^{+8.2}_{-7.6}$ for \bztomutau
 and
 $N^{}_{\rm sig} = 0.3\,^{+8.8}_{-8.2}$ for \bztoetau.
 Both yields are consistent with zero.
 In the \bztomutau sample, we observe $(17\pm 10)$ \bztodpiplus
 events and $(-2\pm 12)$ \bztodstarpi events; these yields are
 consistent with expectations based on MC simulation.


 \section{UPPER LIMIT CALCULATION}

 We calculate upper limits on $N^{}_{\rm sig}$ and
 the branching fractions at 90\% CL\ using a
 frequentist method.
 We first generate sets of MC-simulated events,
 with each set being
 equivalent to the Belle data sample.
 Both signal and background events are generated
 according to their respective PDFs.
 The number of background events
 generated is equal to that obtained from the data fit.
 We vary the number of input signal events, and for each value
 we generate an ensemble of 10\,000 data sets.
 We fit these data sets and calculate the
 fraction ($f_{\rm sig}$) that has a fitted
 signal yield less than that obtained from
 the Belle data (1.8 or 0.3 events).
 Our 90\% CL upper limit on the number of signal
 events ($N_{\rm sig}^{\rm UL}$) is the number of
 input signal events that has $f_{\rm sig}=0.10$.
 We convert $N_{\rm sig}^{\rm UL}$ to an upper limit
 on the branching fraction (${\cal B}^{\rm UL}$) via the formula
  \begin{eqnarray}
  {\cal B}^{\rm UL} & = & \frac{N_{\rm sig}^{\rm UL}}
  {2\times N_{B \overline{B}}\times f^{00} \times \varepsilon }\,.
  \label{eqn:bf}
  \end{eqnarray}
 In this expression,
 $N_{B\overline{B}}$ is the number of $B\overline{B}$ pairs;
 $f^{00}=0.486 \pm 0.006$ is the fraction that
 are $B^0 \overline{B}{}^0$~\cite{pdg}; and
 $\varepsilon$ is the signal efficiency including
 tag-side branching fractions and
 reconstruction efficiencies.

 We include systematic uncertainties
 (discussed below)
 in ${\cal B}^{\rm UL}$ as follows. We divide all systematic
 uncertainties into two types (see Table~\ref{tab:sys}):
 those arising from the numerator of Eq.~\ref{eqn:bf}
 (``additive'' uncertainties), and those arising
 from the denominator of Eq.~\ref{eqn:bf}
 (``multiplicative'' uncertainties). Additive uncertainties arise
 from fitting for the signal yield, while multiplicative
 uncertainties correspond to the number of $B$ decays reconstructed.
 We account for the latter when generating MC data
 sets in our frequentist procedure.
 The number of signal events is varied randomly
 around the nominal input value by the total
 multiplicative uncertainty. Subsequently, after fitting
 an MC data set, we adjust the fitted value $N_{\rm sig}$ by
 a value sampled from a Gaussian distribution
 with mean zero and a width equal to the  total additive
 uncertainty.
 As a final step, to include possible fit bias,
 this value is shifted by an amount obtained by sampling
 a Gaussian distribution with a mean equal to the fit bias
 discussed earlier (the central value) and a width equal
 to the uncertainty in the bias.
 This final value is used when calculating~$f^{}_{\rm sig}$.
 The resulting upper limits for $N^{\rm UL}_{\rm sig}$
 and ${\cal B}^{\rm UL}$ are listed in
 Table~\ref{tab:results}.
 These values are the same as the upper limits expected
 based on MC ($1.6\times 10^{-5}$ for both modes),
 reflecting good agreement between the background levels
 observed in data and the MC.

 \begin{table}[h]
    \caption{Summary of the fit results for $N^{}_{\rm sig}$, and
    the resulting 90\% CL upper limits $N_{\rm sig}^{\rm UL}$
    and ${\cal B}^{\rm UL}$ (see text).}
 \label{tab:results}
  \centering
   \begin{tabular}{l|c c c c c}
   \hline \hline
    Mode    & $\varepsilon$ & $N_{\rm sig}$ & & $N_{\rm sig}^{\rm UL}$ & ${\cal{B}}^{\rm UL}$ \\
           & ($\times 10^{-4}$)  &   & &   & ($\times 10^{-5}$)   \\
   \hline
   \bztomutau & 11.0       & $1.8\,^{+8.2}_{-7.6}$ & & 12.4 & 1.5 \\
   \bztoetau  & 9.8        & $0.3\,^{+8.8}_{-8.2}$ & & 11.6 & 1.6 \\
   \hline \hline
   \end{tabular}
 \end{table}


 \section{SYSTEMATIC UNCERTAINTIES}

 The systematic uncertainties in our measurement
 --\,aside from potential fit bias, which is treated
 separately when setting the upper limits\,-- are
 listed in Table~\ref{tab:sys}.
 Uncertainties in the shapes of the PDFs used for the
 signal are evaluated by varying all fixed parameters by
 $\pm 1 \sigma$;
 the resulting change in the signal yield is taken as
 the systematic uncertainty. The fixed parameters that
 are varied include the correction factors to the shapes
 as obtained from the \bztodpi control samples.
 The fraction of the self-cross-feed signal is fixed to
 the MC value. We vary this fraction by $\pm 50$\% and take
 the resulting change in the signal yield as
 the systematic uncertainty.

 The reconstruction efficiency for $B^{}_{\rm tag}$ is
 evaluated via MC simulation. 
 However, there is uncertainty arising from branching fractions
 for tagging modes that are not well-measured, and from
 unknown decay dynamics of multi-body hadronic decays.
 To account for these effects, a correction factor to
 the reconstruction efficiency is applied. This correction is evaluated
 as done in Ref.~\cite{tagcorr}, by comparing the number of
 events containing both a $B^{}_{\rm tag}$ and a semileptonic
 $B\to D^{(*)}\ell\nu$ decay in data and MC. As the
 branching fractions for $B\to D^{(*)}\ell\nu$ are
 precisely known, and their reconstruction efficiencies
 can be separately calculated, the difference between data
 and MC for $B^{}_{\rm tag}$ reconstruction can be extracted.
 The resulting correction factor is $0.64\pm 0.03$.
 The uncertainty in this value is taken as a systematic uncertainty.
 
 The systematic uncertainty due to charged track reconstruction
 is evaluated using $D^{*+} \to D^0 \pi^+$ decays, with
 $D^0 \to K_S^0\,\pi^+\pi^-$ and $K_S^0 \to \pi^+\pi^-$. The
 resulting uncertainty is 0.35\% per track.
 The uncertainty due to lepton identification is evaluated using
 $e^+e^-\to e^+e^-\gamma^* \gamma^*\to e^+e^-\ell^+ \ell^-$ events.
 The resulting uncertainties are 1.6\% for muons and 1.8\% for
 electrons.

 The systematic uncertainty in the signal reconstruction efficiency
 due to limited MC statistics is $<0.1$\% for both signal modes.
 The systematic uncertainty arising from the number
 of $B\overline{B}$ pairs is 1.4\%, and the known uncertainty on
 $f^{00}$ corresponds to a systematic uncertainty of 1.2\%.

 The total additive (in number of events) and multiplicative
 (in percent) systematic uncertainties are obtained by adding in
 quadrature all systematic uncertainties of that type.

 \begin{table}[h]
  \centering
   \caption{Systematic uncertainties for the branching fraction measurement.
   Those listed in the upper section (``additive'') arise from
   fitting for the signal yield and are listed in number of events;
   those in the lower section (``multiplicative'') arise from
   the number of reconstructed $B$ decays
   and are listed in percent.}
   \label{tab:sys}
   \begin{tabular}{l|c|c}
   \hline \hline
   Source                     & \bztomutau & \bztoetau \\
   \hline
   PDF shapes                 &   0.7      &   0.3     \\
   Self-cross-feed fraction    &  $<0.1$     &   0.1    \\
   \hline
   Total (events)    &   0.7      &   0.3     \\
   \hline
   \multicolumn{3}{c}{}  \\
   \hline
   $B_{\rm tag}$              &   4.5      &   4.5     \\
   Track reconstruction       &   0.3      &   0.3    \\
   Lepton identification      &   1.6      &   1.8     \\
   MC statistics              &    $<0.1$  &  $<0.1$  \\
   Number of $B\overline{B}$ Pairs &   1.4      &   1.4     \\
   $f^{00}$ ($B\overline{B}\to B^0\overline{B}{}^0$ fraction)  &   1.2  &   1.2  \\
  \hline
   Total (\%)  &   5.1      &   5.2     \\
  \hline \hline
  \end{tabular}
 \end{table}


 \section{SUMMARY}

 We have searched for the lepton-flavor-violating decays \bztoltau
 using the full Belle data set. We find no evidence for these
 decays and set the following upper limits on the branching fractions
 at 90\% CL:
 \begin{eqnarray}
  {\cal B}(\mbox{\bztomutau}) & < & 1.5 \times 10^{-5} \\
  {\cal B}(\mbox{\bztoetau})   & < & 1.6 \times 10^{-5} \,.
 \end{eqnarray}
 Our result for \bztomutau is very similar to a recent
 result from LHCb~\cite{btoltauLHCb}.
 Our result for \bztoetau is the most stringent limit to date,
 improving upon the previous limit by
 almost a factor of two. We find no indication of lepton
 flavor violation in these decays.


\section*{ACKNOWLEDGMENTS}

We thank the KEKB group for the excellent operation of the
accelerator; the KEK cryogenics group for the efficient
operation of the solenoid; and the KEK computer group, and the Pacific Northwest National
Laboratory (PNNL) Environmental Molecular Sciences Laboratory (EMSL)
computing group for strong computing support; and the National
Institute of Informatics, and Science Information NETwork 5 (SINET5) for
valuable network support.  We acknowledge support from
the Ministry of Education, Culture, Sports, Science, and
Technology (MEXT) of Japan, the Japan Society for the
Promotion of Science (JSPS), and the Tau-Lepton Physics
Research Center of Nagoya University;
the Australian Research Council including grants
DP180102629, 
DP170102389, 
DP170102204, 
DP150103061, 
FT130100303; 
Austrian Federal Ministry of Education, Science and Research (FWF) and
FWF Austrian Science Fund No.~P~31361-N36;
the National Natural Science Foundation of China under Contracts
No.~11435013,  
No.~11475187,  
No.~11521505,  
No.~11575017,  
No.~11675166,  
No.~11705209;  
Key Research Program of Frontier Sciences, Chinese Academy of Sciences (CAS), Grant No.~QYZDJ-SSW-SLH011; 
the  CAS Center for Excellence in Particle Physics (CCEPP); 
the Shanghai Pujiang Program under Grant No.~18PJ1401000;  
the Shanghai Science and Technology Committee (STCSM) under Grant No.~19ZR1403000; 
the Ministry of Education, Youth and Sports of the Czech
Republic under Contract No.~LTT17020;
Horizon 2020 ERC Advanced Grant No.~884719 and ERC Starting Grant No.~947006 ``InterLeptons'' (European Union);
the Carl Zeiss Foundation, the Deutsche Forschungsgemeinschaft, the
Excellence Cluster Universe, and the VolkswagenStiftung;
L’Institut National de Physique Nucléaire et de Physique des Particules (IN2P3) du CNRS (France);
the Department of Atomic Energy (Project Identification No. RTI 4002) and the Department of Science and Technology of India; 
the Istituto Nazionale di Fisica Nucleare of Italy; 
National Research Foundation (NRF) of Korea Grant
Nos.~2016R1\-D1A1B\-01010135, 2016R1\-D1A1B\-02012900, 2018R1\-A2B\-3003643,
2018R1\-A6A1A\-06024970, 2018R1\-D1A1B\-07047294, 2019K1\-A3A7A\-09033840,
2019R1\-I1A3A\-01058933;
Radiation Science Research Institute, Foreign Large-size Research Facility Application Supporting project, the Global Science Experimental Data Hub Center of the Korea Institute of Science and Technology Information and KREONET/GLORIAD;
the Polish Ministry of Science and Higher Education and
the National Science Center;
the Ministry of Science and Higher Education of the Russian Federation, Agreement 14.W03.31.0026, 
and the HSE University Basic Research Program, Moscow; 
University of Tabuk research grants
S-1440-0321, S-0256-1438, and S-0280-1439 (Saudi Arabia);
the Slovenian Research Agency Grant Nos. J1-9124 and P1-0135;
Ikerbasque, Basque Foundation for Science, Spain;
the Swiss National Science Foundation;
the Ministry of Education and the Ministry of Science and Technology of Taiwan;
and the United States Department of Energy and the National Science Foundation.

\tighten


\begin{thebibliography}{99}
\bibitem{chargeconj}{Throughout this paper, charge-conjugate modes are
         implicitly included unless noted otherwise.}
\end{thebibliography}
\end{document}